\documentstyle[twocolumn]{mn}
\begin{document}
\title[Weak lensing correlations in open and flat universes]
{Weak lensing correlations in open and flat universes}
\author[J.L. Sanz, E. Mart\'\i nez-Gonz\'alez and N. Ben\'\i tez]
{J.L.~Sanz$^1$, E.~Mart\'\i nez-Gonz\'alez$^1$ and N.~Ben\'\i tez
$^{1,2}$\\
$^1$ Instituto de F\'\i sica de Cantabria (CSIC-UC)\\ 
$^2$ Departamento de F\'\i sica Moderna, Universidad de Cantabria\\ 
Facultad de Ciencias, Av. de Los Castros s/n, Santander 39005, SPAIN\\} 
\maketitle
\begin{abstract}
Correlations between the magnification or polarization of background 
sources, induced by gravitational lensing due to the large-scale 
structure, and the positions of foreground galaxies are investigated. 
We found that their amplitude is enhanced with respect to correlations 
for a single population. 
%This means that it should be easier 
%to detect gravitational lensing effects by the large scale structure around 
%foreground galaxy samples which trace these structures than looking in 
%random directions of the sky. 
We analize the dependence of the correlations with the density parameter 
$\Omega$ considering a nonlinear evolution of the matter power spectrum. 
The contribution of the linear evolution is negligible at scales below 
several arcminutes. Detailed results on the dependence of the correlations 
on the redshift of the foreground and background populations for different 
cosmological models are presented. The effect reaches its maximum 
amplitude for foreground populations with relatively small redshifts 
due to the fast increase of the nonlinear matter power spectrum at 
recent times.   
\end{abstract}

\section{Introduction}

The gravitational deflection of photons can be used as a probe 
of the matter distribution along the line of sight to the sources. 
The latter may be at the last scattering surface ($z\simeq 10^3$), 
in the case of the cosmic microwave background (Seljak 1996; 
Mart\'\i nez-Gonz\'alez et al. 1997; Bernardeau 1997), or 
high$-z$ objects as QSOs or galaxies (Blanford et al. 1991; 
Kaiser 1992; Kaiser \& Squires 1993; Bartelmann 1995; Villumsen 1995b; 
Villumsen 1996; Bernardeau, van Waerbeke \& Mellier 1996; Kaiser 1996; 
Jain \& Seljak 1996). Information about the matter fluctuations 
can be obtained on different scales ranging from galaxy haloes to the 
large-scale structure of the universe. 

Many of the theoretical studies on this subject have dealt with 
the polarization or ellipticity produced on background galaxies 
by the large-scale structure of the universe, and there are 
currently several ongoing observational projects trying to detect 
and quantify this effect.
Nevertheless, measuring shear amplitudes as the ones predicted by 
the above mentioned calculations is very difficult 
from a technical point of view (although see Mould et al. 1991; 
Villumsen 1995a), and it is not totally clear if such a precision 
would be routinely achievable in the near future (Ringberg workshop 1997). 

However, there is another observable phenomenon produced 
by gravitational lensing of background sources by foreground mass 
distributions which may have been already detected: QSO-galaxy 
associations due to the magnification bias effect \cite{can81}.  
The surface density of a sample of flux-limited background
sources behind a lens which magnifies them by a factor $\mu$ is changed in 
the form $N'(>S)\propto \mu^{-1}N(S\mu^{-1})$, where 
$N(>S)$ is the unperturbed background source density. If 
$N(>S)\propto S^{-\alpha}$ ( or $N(<m)\propto 10^{0.4\alpha m}$), the 
change in the density can be characterized by the factor 
$q=N'/N=\mu^{\alpha-1}$. Thus, depending on the slope $\alpha$ there may 
be an excess of background sources ($\alpha>1$), a depletion ($\alpha<1$), 
or the 
density may remain unchanged ($\alpha=1$). If we properly choose 
the background source population, so that it has a slope $\alpha$ 
considerably different from 1, there would be a correlation (or 
anticorrelation) between the position of the matter overdensities 
acting as lenses and the background sources. Now, these matter 
perturbations will be traced, up to a bias factor, by galaxies and 
thus, there will be a correlation between these foreground galaxies 
(or any other tracers of dark matter) and the background sources. 

There are several reported associations between foreground 
galaxies and high redshift, background AGNs (see Schneider, 
Ehlers \& Falco (1992); Narayan \& Bartelmann (1996) or 
Wu (1996) for reviews), but only a few of these studies extend 
to relatively large scales. Bartelmann \& Schneider (1994) found a strong
association between galaxies selected from the IRAS Faint Source Catalogue 
and high$-z$ AGN from the 1Jy catalogue. In Ben\'\i tez \& Mart\'\i nez-
Gonz\'alez (1995) it was found that red APM galaxies tracing large scale 
structures were correlated with 1Jy QSOs. Another sample of radio loud QSOs, 
extracted from the PKS catalogue has also been found to be correlated with 
COSMOS galaxies \cite{ben97}, with a correlation scale of several arcmin. 
Other studies considering the correlation between galaxy clusters 
and high-z QSOs (Seitz \& Schneider 1995, Wu \& Han 1996) have also 
found positive results.

In this paper, we shall study the effects of weak gravitational lensing 
by foreground matter fluctuations on a population of background sources at 
high$-z$. We consider different values of $\Omega$ and model the 
fluctuations assuming CDM with a power spectrum whose evolution in 
time follows a standard ansatz (Hamilton et al. 1991, 
Peacock \& Dodds 1996, linear and non-linear contributions are considered).
%A semirealistic approximation to a population of foreground objects (usually
%galaxies) producing the lensing can be done through a global bias parameter
%$b$.
We assume that these matter perturbations are traced, up to a global bias 
parameter $b$ by galaxies. More specifically, we shall explore the behavior 
of $C_{p\delta}$, i.e. the large-scale correlation between the ellipticity 
of background galaxies and the position of foreground ones, which apparently 
has not been considered in the literature. We shall also consider in detail 
other correlations (in particular their dependence on $\Omega$) such as 
$C_{\mu\delta}, C_{\mu\mu}$ i.e. magnification-foreground galaxies and 
magnification-magnification. $C_{\mu\mu}$ can be indirectly estimated 
through the galaxy-galaxy correlation function (Villumsen 1995b). 
However, measuring $C_{\mu\delta}$ offers several advantages over 
$C_{\mu\mu}$ from the observational point of view. In the first place, 
$C_{\mu\delta}$ has an amplitude several times higher than $C_{\mu\mu}$.
Besides, if the foreground and background galaxy populations are 
properly selected so that there is no redshift overlap between them 
(e.g high$-z$ QSOs and bright galaxies), one does not have to bother 
about intrinsic correlations: any measured effect should be caused by 
gravitational lensing. 

Section 2 develops the formalism dealing with weak gravitational lensing for a
flat and open cosmological model, the concepts of magnification and 
polarization (or ellipticity) and the different correlations. In 
section 3 appear the main theoretical results as well as comments 
on different observational perspectives. Finally, in section 4 we give 
the conclusions of the paper.

\section{Formalism}\label{form}
 
\subsection{Geodesics in a perturbed Friedmann universe}

We will consider the propagation of photons from a source at redshift $z$ to 
the observer ($z = 0$), the universe being a perturbed Friedmann model with
vanishing pressure.  For scalar perturbations, the metric in the conformal 
Newtonian gauge is given in terms of the scale factor $a(\tau )$ and a single 
potential $\phi (\tau, \bmath{x})$, that satisfies the Poisson equation, as 
follows (Mart\'\i nez-Gonz\'alez et al. 1997)
\begin{equation}
ds^2 = a^2(\tau )[-(1 + 2\phi )d{\tau }^2 + (1 - 2\phi ){\gamma }^{-2}
{\delta }_{ij}dx^idx^j],
\end{equation}
\[
 \gamma = 1 + \frac{k}{4}{\bmath{x}}^2, 
\]
\noindent we take units such that $c = 8\pi G = a_o = 2H_o^{-1} = 1$ and 
$k/(4\mid 1-\Omega\mid) = 0, -1 , +1$ denote the flat, open and closed 
Friedmann background universe.

\noindent Assuming a perturbation scheme ("weak lensing"), the null geodesic 
equation for the previous metric can be integrated in the form
$\bmath{x} = \lambda \bmath{n} + \bmath{\epsilon}$, where $\bmath{n}$ is 
the direction of observation and $\lambda$ is the distance to the photon 
in the background metric, i.e.
 
\begin{eqnarray}
&\lambda =&{\tau}_o - \tau ~~~~(k = 0),\nonumber\\
&\lambda =&{(1 - \Omega )}^{-1}{\rm{tanh}}
[(1 - \Omega )( {\tau}_o - \tau )]~~~~ (k = -1). 
\end{eqnarray}

\noindent The perturbation $\bmath{\epsilon}$ can be decomposed 
in a term parallel to the direction of observation $\bmath{n}$ 
and a term, ${\bmath{\alpha }}_{\bot}$, ortoghonal to such a 
direction. The last term is given by 

\begin{equation}
\bmath{\alpha}_{\bot} = 2\int_0^{\lambda}d{\lambda}'
W(\lambda , {\lambda}')\bmath{\nabla}_{\bot}\phi ({\lambda}', \bmath{x} =
{\lambda}'\bmath{n})\ \ \
\end{equation}
 
\noindent where $W(\lambda, {\lambda}')$ is a window function
 
\begin{equation}
a(\lambda) = \frac{(1 - \lambda )^2}{1 + k{\lambda}^2/4},\ \ \ 
W(\lambda, {\lambda}') = (\lambda - {\lambda}')
\frac{1 + k\lambda {\lambda}'/4}{1 + k{{\lambda}'}^2/4}.
\end{equation}
 
For photons that are propagated from a source at redshift $z$ (distance
$\lambda $) to the observer($z_o = 0$ or ${\lambda}_o = 0$), the 
lensing vector $\bmath{\beta}$ is defined in the usual way: 
\[
\bmath{\beta}\equiv\bmath{n} - \frac{\bmath{x} - \bmath{x}_o}
{|\bmath{x} - \bmath{x}|}
\]
Thus we find
\begin{equation}
\bmath{\beta} = \frac{1}{{\lambda}}\bmath{\alpha}_{\bot}({\lambda}).
\end{equation}
Once we have obtained the expression for the trajectory of the photon in the
conformal Newtonian gauge, it is easy to calculate everything in the conformal
synchronous-comoving gauge (Mart\'\i nez-Gonz\'alez et al. 1997). The lensing
vector in such a gauge (that is the appropriate one from the point of view of
observations) is given by the expression (5) plus some additional terms that
can be interpreted as Doppler contributions at the source and observer and an
acceleration term at the observer. The last two terms can be estimated from the
Doppler velocity respect to the cosmic microwave background and from our local
infall towards the Virgo cluster (or Great Attractor). These extra
contributions are very small, so the lensing vector $\bmath{\beta}$ in the
synchronous-comoving gauge is approximately given by $\bmath{\beta}$, 
as defined by equations (3, 5). 

\subsection{Magnification and Polarization}

Let us assume a population of 
background sources (e.g. quasars or galaxies), placed at 
different distances $\lambda $
with a distribution $R_b(\lambda )$ ($\int_0^1d\lambda R_b(\lambda) = 1$). 
Then, we can define the integrated lensing vector $\bmath{\beta}(\bmath{n}) =
\int_0^1d\lambda R_b(\lambda)\bmath{\beta}(\lambda ,\bmath{n})$ and taking into
account equations (3-5) we obtain

\begin{equation}
\bmath{\beta}(\bmath{n}) = \bmath{D}S,~~~~~ D^i\equiv ({\delta}^{ij}-
n^in^j)\frac{\partial}{\partial n^j},
\end{equation}
\[
S(\bmath{n})\equiv 2\int_0^1d\lambda T_b(\lambda)\phi (\lambda , \bmath{x} =
{\lambda}\bmath{n}),
\]
\begin{equation}
T_b(\lambda)\equiv \frac{1}{\lambda}\int_{\lambda}^1d{\lambda}'\frac{1}
{{\lambda}'}R_b({\lambda}')W({\lambda}', \lambda ).
\end{equation}

\noindent If we take the derivative along the plane orthogonal to the 
direction of
observation $\bmath{n}$, we get the convergence tensor ${\theta}_{ij}\equiv
D_j{\beta}_i = D_jD_iS$ which can be decomposed in the form

\begin{eqnarray}
\theta_{ij} &=  \kappa h_{ij} + p_{ij}, 
&h_{ij}\equiv {\delta}_{ij}- n_in_j,\nonumber\\
\kappa (\bmath{n}) &= \frac{1}{2} h^{ij}{\theta}_{ij},
&p_{ij}(\bmath{n}) = {\theta}_{ij} - \kappa h_{ij}
\end{eqnarray}

For weak lensing, the convergence scalar $\kappa (\bmath{n})$ is related 
to the magnification by $\mu  (\bmath{n}) = 1+2\kappa (\bmath{n})$. 
Moreover, the polarization components in the plane orthogonal to 
$\bmath{n}$ can be defined in the standard way $p_1\equiv p_{11} - 
p_{22}, p_2\equiv 2p_{12}$ and the complex polarization is 
$p \equiv p_1 + ip_2$. So, we can define the scalar polarization 
$p^2\equiv pp* =2p_{ij}p^{ij}$.

\subsection{Background-Foreground correlations}

Let us consider a second population of foreground sources (e.g. galaxies) 
placed at different distances $\lambda$ with a distribution 
$R_f(\lambda )$ ($\int_0^1d\lambda R_f(\lambda) = 1$). 
If $\delta (\lambda ,\bmath{x})$ represents the density fluctuation 
(that satisfies the Poisson equation 
$({\nabla }^2 + 3k)\phi =\frac{1}{2}{\rho}_ba^2\delta $), 
we shall assume that there is a constant bias factor $b$ relating
the number fluctuation, $\frac{\delta N}{N}$, and such an overdensity. Then,
we can define the integrated overdensity ${\delta}(\bmath{n}) =
\int_0^1d\lambda R_f(\lambda){\delta}(\lambda ,\bmath{n})$. 

Taking into account equations (3-7), we obtain

\begin{equation}
\langle {\beta}_i(\bmath{n}) {\delta}(\bmath{n})\rangle = 
D_i\langle S(\bmath{n}) {\delta}(\bmath{n})\rangle , 
\end{equation}
\begin{equation}
\langle S(\bmath{n}) {\delta}(\bmath{n})\rangle = 2\int_0^1d\lambda 
T_b(\lambda)\int_0^1d{\lambda}' R_f({\lambda}')C_{\phi \delta}(\lambda
,{\lambda}'; r),
\end{equation}

\noindent where $C_{\phi \delta}(\lambda ,{\lambda}'; r)
\equiv \langle \phi (\lambda ,\bmath{x})\delta ({\lambda}',\bmath{x}')\rangle 
, r\equiv {\mid \bmath{x}- \bmath{x}'\mid}_{\Omega}$. Using the Limber 
approximation (see Appendix), i. e. only a small region $r$ is 
contributing with ${\lambda}' \simeq \lambda$, the previous equation can be 
approximated by

\begin{eqnarray}
\langle S(\bmath{n}) {\delta}(\bmath{n'})\rangle &\simeq& 
4{\int}_0^1d\lambda T_b(\lambda) 
R_f(\lambda)[1 - (1 - \Omega ){\lambda}^2]\times\nonumber\\ 
&\times&{\int}_{\theta s}^{\infty}dr
\frac{C_{\phi \delta}(\lambda ; r)}
{{(r^2 - {\theta}^2 s^2)}^{1/2}},
\end{eqnarray}

\noindent where now appears the correlation at a single time and $s$ is given 
by $s\equiv\lambda[1-(1-\Omega)\lambda^2]^{-1}$. Introducing the 
power spectrum $P(a, k)$ of the matter density fluctuations defined by

\begin{equation}
\langle {\delta}_k(a){\delta}^*_{k'}(a)\rangle \equiv P(a, k)
{\delta}^3(\bmath{k} - \bmath{k}').
\end{equation}

\noindent and the relation $P_{\phi ,\delta }(\lambda , k)=-6\Omega 
P(\lambda , k)a^{-1}(\lambda)$, obtained via the Poisson equation 
for small scales, the last expression (11) is

\begin{eqnarray}
\langle S(\bmath{n}) {\delta}(\bmath{n})\rangle &\simeq& 
-\frac{6\Omega}{\pi}\int_0^1d\lambda \frac{ T_b(\lambda)R_f(\lambda)\lambda^2}
{s^2(1 - \lambda )^2}\times\nonumber\\
&\times&\int_0^{\infty}dkk^{-1}P(\lambda , k)J_0(ks\theta )
\end{eqnarray}

Finally, taking another $D_i$ derivative on the equation (9), we get the
following correlations

\begin{equation}
C_{\mu\delta}(\theta )\equiv 2\langle \kappa 
(\bmath{n})\delta(\bmath{n}')\rangle = A_0, 
\end{equation}
\begin{equation}
C_{p\delta}(\theta )\equiv \langle p(\bmath{n})\delta(\bmath{n}')\rangle 
= A_2,
\end{equation}
\begin{eqnarray}
A_i&\equiv& \frac{6\Omega}{\pi}\int_0^1d\lambda T_b(\lambda) R_f(\lambda)
(\frac{\lambda}{1 - \lambda })^2\times\nonumber\\
&\times&\int_0^{\infty}dkkP(\lambda , k)
J_i(ks\theta )\label{ai}
\end{eqnarray}
\noindent and $J_i$ is the Bessel function of 1st kind. These are the basic
formulas to be applied when two different populations are correlated. In
particular, the background-foreground correlation function is given by 
Bartelmann (1995).

\begin{equation}
{\xi}_{bf}(\theta ) = b(\alpha - 1)C_{\mu ,\delta },
\end{equation}

\noindent where $\alpha $ is the slope of the background source number 
counts.

If the background population is concentrated at a certain redshift $z_b$ 
corresponding to a distance ${\lambda}_b$, having a Dirac delta 
distribution $R_b=\delta_D(\lambda-\lambda_b)$, then the window 
$T_b$ is given by
\[
T_b(\lambda)  =  0, \lambda \geq {\lambda}_b\
\]
\begin{equation}
T_b(\lambda)  =  (\frac{1}{\lambda} - \frac{1}{\lambda}_b)
\frac{1 - (1 - \Omega )\lambda {\lambda}_b}{1 - (1 - \Omega ){\lambda }^2}, 
\lambda \leq {\lambda}_b
\end{equation}
If the foreground population has also a Dirac delta distribution form 
$R_f=\delta_D(\lambda-\lambda_f)$, then the window $T_b$ in equation (16) 
is $T_b=T_b(\lambda_f)$ and 
\begin{eqnarray}
A_i&=&
\frac{6\Omega\lambda_f (\lambda_b-\lambda_f)[1-(1-\Omega)\lambda_f\lambda_b]}
{\pi\lambda_b(1-\lambda_f)^2[1-(1-\Omega)\lambda_f^2]}\times\nonumber\\
&\times&\int_0^\infty dk k P(\lambda_f,k)J_i(k\theta s_f)
\end{eqnarray}
where $s_f=\lambda_f[1-(1-\Omega)\lambda_f^2]^{-1}$.

\subsection{Background autocorrelations}
 
Let us consider a population of background sources (e.g. galaxies) 
placed at different
distances $\lambda$ with a distribution $R_b(\lambda )$ ($\int_0^1d\lambda
R_b(\lambda) = 1$). We are now interested in the gravitational lensing
properties induced by the population itself.
The magnification-magnification $C_{\mu \mu }$, polarization-polarization 
$C_{pp}$ and magnification-overdensity $C_{\mu \delta}$ correlations are 
defined by

\begin{equation}
C_{\mu \mu }(\theta )\equiv 4\langle \kappa (\bmath{n})\kappa (\bmath{n}')
\rangle = 
D^iD_iD^{j'}D_{j'}\langle S(\bmath{n})S(\bmath{n}')\rangle,
\end{equation}
\begin{eqnarray}
C_{pp}(\theta )&\equiv& 2\langle p^{ij}(\bmath{n})p_{ij}(\bmath{n}')
\rangle \equiv\langle p(\bmath{n})p^{*}(\bmath{n}')\rangle=\nonumber\\
& =&
D^iD_{i'}D^jD_{j'}\langle S(\bmath{n})S(\bmath{n}')\rangle - 
\frac{1}{2}C_{\mu \mu }(\theta ),
\end{eqnarray}
\begin{equation}
C_{\mu \delta}(\theta )\equiv 2\langle \kappa (\bmath{n})\delta (\bmath{n}')\rangle =
D^iD_i\langle S(\bmath{n})\delta(\bmath{n}')\rangle.
\end{equation}

\noindent After an straightforward calculation, one obtains

\begin{eqnarray}
\langle S(\bmath{n})S(\bmath{n}')\rangle &=&
\frac{72{\Omega}^2}{\pi}\int_0^1d\lambda T^2_b(\lambda)
\frac{{[1 - (1 - \Omega ){\lambda}^2]}^3}{{(1 - \lambda )}^4}\times\nonumber\\
&\times&\int_0^1dkk^{-3}P(\lambda ,k)J_0(ks\theta ),
\end{eqnarray}

\noindent where the Limber approximation has been assumed and $T_b$ is given by
equation (7). Now, taking into
account the definitions for the correlations, one gets (see also Kaiser 1992,
1996; Jain \& Seljak 1996)
\begin{eqnarray}
C_{\mu \mu }(\theta ) &=&
\frac{72\Omega^2}{\pi}\int_0^1d\lambda \frac{ T^2_b(\lambda)\lambda^4}
{(1 -\lambda )^4[1 - (1 - \Omega )\lambda^2]}\times\nonumber\\
&\times&\int_0^\infty dkkP(\lambda ,k)J_0(ks\theta ),
\end{eqnarray}
\begin{equation}
 C_{pp}(\theta ) = C_{\mu\mu}(\theta ),
\end{equation}

\noindent the last expression in agreement with Villumsen (1995b). 
For a population with a Dirac delta distribution 
$R_b=\delta(\lambda-\lambda_b)$, the window $T_b(\lambda)$ 
is given by equation (18). 

Moreover, $C_{\mu \delta }(\theta )$ is given by equation (14) but now $R_f$
must be substituted by $R_b$. Trivially, for a strongly peaked $z$-distribution
such a correlation is very small (see the dependence on ${\lambda}_b -
{\lambda}_f$ in equation (19)).

It is interesting to note that the observed correlation function $w(\theta)$
associated to a population is given by the following contributions (Villumsen
1995b)

\begin{equation}
\frac{\delta N}{N}(\bmath{n}) = b\delta (\bmath{n}) + 
(\alpha - 1){\mu}(\bmath{n}),
\end{equation}

\begin{eqnarray}
\omega(\theta) &=& \left\langle \frac{\delta N}{N}(\bmath{n})
\frac{\delta N}{N}(\bmath{n}')\right\rangle =\nonumber\\ 
&=&b^2C_{\delta \delta } + 2b(\alpha - 1)C_{\mu \delta} 
+ {(\alpha - 1)}^2C_{\mu \mu},
\end{eqnarray}

\noindent where $C_{\delta \delta }$ is the matter correlation function. So,
gravitational lensing modifies the intrinsic correlation approximately by the
magnification term.

\section{Results}
\begin{figure}
\vspace{6cm}
\caption{$C_{\mu \delta}(\theta)$ for a foreground population peaked at
$z_f=0.2$ and a background one at $z_b=1$ and for three values of $\Omega$: 
1 (solid), 0.3 (dotted) and 0.1 (dashed). The three bottom lines represent 
the linear contribution for the same $\Omega$ values.}
\end{figure}

With the formalism presented in the previous section we have calculated the
correlations $C_{\mu\delta}$, $C_{p\delta}$ and $C_{\mu\mu}$. 
We assume a CDM model with a primordial
Harrison-Zeldovich spectrum, a Hubble parameter h$=0.5$ (H$=100$h km s$^{-1}$
Mpc$^{-1}$) and flat as well as open universe models. For the power spectrum 
we have used the fit given by equation (G3) of Bardeen et al. (1986) which is
normalized to the cluster abundance: $\sigma_8=0.6\Omega^{-F(\Omega)},\
F(\Omega)\equiv 0.34+0.28\Omega-0.13\Omega^2$,  following Viana and Liddle 
(1996) (see also 
White, Efstathiou and Frenk 1993; Eke, Cole and Frenk 1996; see 
Kaiser 1997 for a discussion on alternative normalizations). 
For the nonlinear evolution of the power spectrum we use the 
recently improved fitting formula given by Peacock and Dodds (1996). 
That formula is based on the Hamilton et al. (1991) scaling procedure 
to describe the transition between linear and nonlinear regimes. It 
accounts for the correction introduced by Jain, Mo and White (1995) 
for spectra with $n\la -1$ and applies to flat as well as open universes.
\begin{figure}
\vspace{12cm}
\caption{$\bar C_{\mu \delta}(1')$ as a function of the foreground 
population redshift $z_f$ and for background redshifts $z_b=0.5,1,2,5$. (a)
$\Omega=1$, (b) $\Omega=0.3$, (c) $\Omega=0.1$.}
\end{figure}

For the redshift distributions of the background and foreground sources we
consider a Dirac delta distribution peaked at $\lambda_b$ and $\lambda_f$
respectively, $R_b(\lambda)=\delta_D(\lambda-\lambda_b)$ and 
$R_f(\lambda)=\delta_D(\lambda-\lambda_f)$. These simple distributions 
are very useful since they reduce the calculations and the results differ 
only slightly when compared to other more realistic distributions (see below). 
$C_{\mu\delta}$ and $C_{p\delta}$ are computed using equations (14,15,19).
$C_{\mu\mu}=C_{pp}$ is obtained from equation (25) where the function $T_b$ 
is given by equation (18).

%Equation (16) for 
%$C_{\mu\delta}$ and $C_{p\delta}$ now simplifies to
%$C_{\mu\mu}$ is computed from equation (23) where the function $T_b$ now
%becomes
%\begin{equation}
%T_b(\lambda)=\frac{\lambda_b-\lambda}{\lambda\lambda_b}\frac{1-(1-\Omega)
%\lambda\lambda_b}{1-(1-\Omega)\lambda^2}\ \ , \lambda<\lambda_b
%\end{equation}
%and $T_b(\lambda)=0$ for $\lambda>\lambda_b$.

\subsection{Background magnification-foreground matter crosscorrelations}

The crosscorrelation $C_{\mu\delta}(\theta)$ for a population of background 
sources 
peaked at $z_b=1$ and another of foreground lenses at $z_f=0.2$ is given in
figure (1). The effect is maximum at zero lag and rapidly decreases for scales
above a few arcmin. The amplitude is always above a few percent for scales
$\la 1'$ and at these scales the linear contribution is negligible 
compared to the nonlinear one for all the $\Omega$ values. Notice that 
$C_{\mu\delta}$ increases when $\Omega$ decreases whereas considering only the
linear contribution the situation is reversed.

From the observational point of view it is useful to calculate the average
crosscorrelation (or equivalently the mean relative excess of background
sources around foreground lenses) within a given radius $\theta$, 
$\bar C_{\mu\delta}(\theta)$.
The variation of $\bar C_{\mu\delta}(1')$ with $z_f$ for different values of 
$z_b$ and a $1'$ radius is shown in figures (2a,b,c). A maximum amplitude of a
few percent is obtained at a $z_f$ in the
range $0.1-0.25$ for all background populations and all cosmological models.
(At this point it is important to recall that to compare with observations of 
background-foreground object
correlations $\bar \xi_{bf}(\theta)$, the bias factor $b$ of the foreground 
population and the slope $\alpha$ of the background population enter in the
calculation following equation 16).
It is interesting to notice that there are already available large galaxy
samples, like the APM or COSMOS catalogues, which peak at a redshift within 
that range (for the APM catalogue $<z>=0.16$ for a magnitude limit $B_J=20.5$,
see Efstathiou 1995). Moreover, the use of realistic redshift distributions to
represent the foreground and background populations (with a bell-like shape 
similar to that of the APM one), changes the results in only $\approx 10\%$
for the relevant angular scales
compared to the Dirac delta distribution used here. 
Those catalogues, which except for a bias factor are assumed to follow
the large scale matter distribution, are therefore very suitable to
crosscorrelate with a background source population. This has already been done
by Ben\'\i tez and Mart\'\i nez-Gonz\'alez (1995, 1997) for the 1 Jansky and
Parkes samples of radio loud QSOs as background populations, finding clear 
evidences of positive crosscorrelations. In figure 3, we show
$C_{\mu\delta}(\theta)$ for $0.1\leq \Omega \leq 1$ and for $z_f=0.15$ and
$z_b=1$, mean redshift values appropriate for the galaxy and radio QSO samples 
considered above. The dependence of $C_{\mu\delta}$ with $\Omega$ is more
relevant at small angular scales. In figure 4, we also represent the average
crosscorrelation $\bar C_{\mu\delta}(1')$ as a function of $\Omega$. A rough
comparison between its expected 
amplitude and the measured value for the COSMOS-Parkes samples, as given in
figure 4 of Ben\'\i tez and Mart\'\i nez-Gonz\'alez 1997, shows agreement for
realistic values of $\Omega$ and the bias parameter. 
A detailed comparison of the 
theoretical calculations with the observational results will be given 
elsewhere. Dolag and Bartelmann (1997) have 
recently presented calculations of the 
QSO-galaxy correlation function for such QSO and galaxy populations 
produced by gravitational lensing due to the Large Scale Structure, 
following a similar theoretical scheme.   
\begin{figure}
\vspace{6cm}
\caption{$C_{\mu \delta}(\theta,\Omega)$ for a foreground population peaked at
$z_f=0.15$ and a background one at $z_b=1$.} 
\end{figure}

\begin{figure}
\vspace{6cm}
\caption{$\bar C_{\mu\delta}(1')$ as a function of $\Omega$ for the same 
values of $z_f$ and $z_b$ as in figure (3).
}
\end{figure}
 
\subsection{Background polarization-foreground matter crosscorrelations}
\begin{figure}
\vspace{6cm}
\caption
{$C_{p \delta}(\theta)$ for a foreground population peaked at
$z_f=0.3$ and a background one at $z_b=1$ and for three values of $\Omega$: 
1 (solid), 0.3 (dotted) and 0.1 (dashed). The three bottom lines represent 
the linear contribution for the same $\Omega$ values.}
\end{figure}

The crosscorrelation between the polarization of a background source
population peaked at $z_b$ and the matter density fluctuations peaked at
$z_f$, $C_{p\delta}(\theta)$, is given in figure (5) for $z_b=1$ and
$z_f=0.3$. The maximum value is in the angular range $0.4'-1'$ and this scale
is smaller for low $\Omega$ models. As in the case of $C_{\mu\delta}$, the
linear contribution is negligible; however, it peaks at a much larger angle
of $\sim 10'$ as a consequence of the much larger scales which contribute to
the linear level. Including the nonlinear evolution, a correlation of $\sim
1\%$ can be expected at angular scales $\la 1'$ for realistic models of
structure formation. The use of realistic redshift distributions to
represent the foreground and background populations (with a bell-like 
shape typical of magnitude limited samples) changes the results in 
only $\approx 10\%$ for the relevant angular scales
compared to the Dirac delta distribution used here.

In analogy to the previous subsection, it is useful to calculate the average
crosscorrelation within a given radius $\theta$, $\bar C_{p\delta}(\theta)$.
The variation of $\bar C_{p\delta}(1')$ with $z_f$ for several values of 
$z_b$ and $\Omega$ is shown in figures (6a,b,c). Maximum amplitudes of the 
order of $1\%$ are found for $z_f$ within a relatively wide range $0.2-0.5$.
The amplitude grows appreciably with $z_b$ being a factor of $\approx 2$ 
difference between $z_b=0.5$ and $2$. Considering the behavior of 
$\bar C_{p\delta}(\theta)$ from these figures, suitable populations 
to detect the crosscorrelation would be a foreground sample with $z_f<1$ 
and a background one peaked at $z_b\ga 2$.

\begin{figure}
\vspace{12cm}
\caption
{$\bar C_{p \delta}(1')$ as a function of the foreground 
population redshitf $z_f$ and for background redshifts $z_b=0.5,1,2,5$. (a)
$\Omega=1$, (b) $\Omega=0.3$, (c) $\Omega=0.1$.}
\end{figure}

\subsection{Magnification and Polarization autocorrelations}

In this subsection we concentrate on a single population of background 
sources peaked at a given $z_b$ and calculate $C_{\mu\mu}(\theta)$. 
This is done in figure (7) for $z_b=1$. The maximum effect is at zero 
lag and its amplitude is relatively small $<1\%$. The nonlinear contribution 
clearly dominates over the linear one but now the amplitude 
grows with $\Omega$, contrary to the crosscorrelations 
$C_{\mu\delta}(\theta)$ and $C_{p\delta}(\theta)$. Our results are in 
agreement with Jain \& Seljak (1996), and with Kaiser (1997) 
for one degree scales (see also those papers for a more detailed analysis 
of $C_{\mu\mu}$), where the dominant contribution is the linear one.
Considering also the slope of the population of sources 
following equation (26), $C_{\mu\mu}$ 
could be estimated from the observed autocorrelation of faint galaxies. 
Nevertheless, measuring $C_{pp}$ should be more feasible from the practical 
point of view, as we do not have to disentangle the contribution caused by 
lensing from the intrinsic correlations; it is usually assumed 
(and strongly hoped) that the intrinsic ellipticities of background 
galaxies are not correlated.  

\begin{figure}
\vspace{6cm}
\caption{$C_{\mu \mu}(\theta)$ (or equivalently $C_{pp}(\theta)$) for a 
population peaked at $z_b=1$ and for three values of $\Omega$: 
1 (solid), 0.3 (dotted) and 0.1 (dashed). The three bottom lines represent 
the linear contribution for the same $\Omega$ values.}
\end{figure}
\section{Conclusions}
We have obtained the expressions for the correlations between the 
magnification or polarization of background sources and the foreground matter
distribution as function of the nonlinear evolution of the power spectrum. 
These formuli are valid for flat and open universes. 

For the crosscorrelation of the background magnification and foreground matter
distribution, $C_{\mu\delta}(\theta)$, the maximum is at zero lag and the
amplitude remains above a few percent for scales $\la 1'$. 
$C_{\mu\delta}(\theta)$ increases significantly when $\Omega$ decreases. The
linear contribution is negligible compared to the nonlinear one for the
relevant scales below a few arcmin. Varying the redshift of the foreground
population, $z_f$, a maximum amplitude of a few percent for the integrated
correlation $\bar C_{\mu\delta}(1')$ is obtained at $0.1\la 
z_f\la 0.25$ for all background populations and all cosmological models.

The crosscorrelation of the background polarization and foreground matter
distribution, $C_{p\delta}(\theta)$, presents a maximum of the order of 
$1\%$ at a non-null angle, typically in the range $\approx 0.4'-1'$. 
Fixing the redshifts $z_f$ and $z_b$ of the two populations, the angular 
scale of the maximum decreases with $\Omega$.
$C_{p\delta}(\theta)$ increases significantly when $\Omega$ decreases. 
The linear contribution is negligible compared to the nonlinear one for the
relevant scales below a few arcmin. Varying the redshift of the foreground
population, $z_f$, a maximum amplitude of $\sim 1\%$ for the integrated
correlation $\bar C_{p\delta}(1')$ is obtained within a relatively wide range
$0.2\la z_f\la 0.5$. This amplitude grows appreciably with $z_b$
being a factor of $\approx 2$ difference between $z_b=0.5$ and $2$. 

Finally, the correlation of the magnifications for a single population,
$C_{\mu\mu}(\theta)$, has a relatively small maximum amplitude $\la 1\%$ 
in all cases. The amplitude grows with $\Omega$, contrary to the
crosscorrelations $C_{\mu\delta}(\theta)$, $C_{p\delta}(\theta)$.

\section*{Acknowledgements}
The authors thank Matthias Bartelmann, Klaus Dolag and Peter Schneider 
for interesting and useful discussions. We also thank Jos\'e Revuelta 
for his help with the aesthetic aspects of this paper. JLS, EMG and NBL 
acknowledge financial support from the Spanish DGES, project 
PB95-0041. NB acknowledges a Spanish M.E.C. Ph.D. scholarship.  

\section*{Appendix}
Some equations (e.g. equation (11)) have been obtained assuming the Limber
approximation, i.e. one assumes that the maximum scale of appreciable
correlation is small compared to the typical distances to the foreground and
background objects in the samples. So, if we consider a double integral with
the same form as equation (10), the contribution to the integral is appreciable
only when the points are nearly at the same time $\lambda \simeq {\lambda}'$
and their separation is small, 
${\mid \bmath{x}-\bmath{x}'\mid}_{\Omega}\ll \lambda$, When $\theta \ll 1$,
the spatial separation of two points is 

\begin{eqnarray}
r^2 &=& s^2 + {s'}^2 -2ss'\cos \theta 
\simeq (s - s')^2 + ss'{\theta}^2,\nonumber\\  
s&\equiv& \lambda[1-(1-\Omega)\lambda^2]^{-1}.
\end{eqnarray}

\noindent Defining the new variable $t=\lambda'-\lambda$, 
the separation
can be rewritten as 
$r\simeq[t^2 +\lambda^2\theta^2]^{1/2}[1-(1-\Omega)\lambda^2]^{-1}$.

Therefore, the double integral can be approximated by

\begin{eqnarray}
I&\equiv& \int_0^1d{\lambda}' T_b({\lambda}')\int_0^1d{\lambda} 
R_f({\lambda})C(\lambda,{\lambda}'; r) \simeq\nonumber\\
&\simeq&\int_0^1d{\lambda'}R_f({\lambda'})T_b({\lambda'})
\int_{-\lambda'}^{1-\lambda'}dt C(\lambda'; r(t)),
\end{eqnarray}

\noindent where $C(\lambda'; r(t))\equiv C(\lambda', \lambda'; r(t))$ and 
$T_b(\lambda +t)\simeq T_b(\lambda)$, the latter being a smooth function. The 
previous integral for
the variable $t$ can be approximated by $2\int_0^{\infty}$  if one assumes that
the foreground sources are placed at distances far away from the observer and
Hubble distance (case of practical interest). Finally, changing the variable
$t$ by $r$ one gets equation (11). We remark that equation (11) is also valid
when a Dirac distribution is assumed to represent the foreground redshift 
distribution.
The Limber approximation in Fourier space has been also considered by Kaiser
(1992) in the context of weak gravitational lensing.


\begin{thebibliography}{}

\bibitem[Bardeen et al. 1986]{bardeen86}Bardeen, J.M., Bond, J.R., Kaiser, N.
\& Szalay, A.S. 1986, Apj 304, 15

\bibitem[Bartelmann 1995]{bar95}Bartelmann, M. 1995, A\&A, 298, 661

\bibitem[Bartelmann \& Schneider 1994]{bar94} Bartelmann, M., \&
Schneider, P. 1994, A\&A, 284, 1

\bibitem[Ben\'\i tez \& Mart\'\i nez-Gonz\'alez  1995]{ben95}  
Ben\'\i tez, N., \& Mart\'\i nez-Gonz\'alez, E., 1995, ApJL, 339, 53

\bibitem[Ben\'\i tez \& Mart\'\i nez-Gonz\'alez 1997]{ben97}  
Ben\'\i tez, N., \& Mart\'\i nez-Gonz\'alez, E., 1997, ApJ 447, 27

\bibitem[Bernardeau 1996]{ber96}Bernardeau, F. 1996, astro-ph/9611012

\bibitem[Bernardeau, van Waerbeke \& Mellier 1996]{bvm96}Bernardeau, F., 
van Waerbeke, L. \& Mellier, Y., 1996, astro-ph/9609122

\bibitem[Blanford et al. 1991]{bla91}Blanford, R., Saust, A. B., 
Brainerd, T. G. and Villumsen, J. V. 1991, MNRAS, 251, 600

\bibitem[Canizares 1981]{can81} Canizares, C.R., 1981, Nature, 291, 620

\bibitem[Dolag \& Bartelmann 1997]{dol97}Dolag, K., \& Bartelmann, M., 1997, 
submitted to MNRAS, astro-ph/9704217.

\bibitem[Efstathiou 1995]{efs95} Efstathiou, G. 1995, MNRAS, 272, L25

\bibitem[Eke et al.  1996]{}Eke, V.R., Cole, S. \& Frenk, C.S. 1996, MNRAS,
282, 263.

\bibitem[Hamilton et al 1991]{hamilton91} Hamilton, A.J.S., Kumar, P., Lu, E.
\& Matthews, A. 1991, ApJL 374, L1.

\bibitem[Jain \& Seljak 1996]{jai96}Jain, B. \& Seljak, U., 
1996, astro-ph/9611077

\bibitem[Jain, Mo \& White 1995]{jai95}Jain, B., Mo, H.J. \& White, 
S.D.M. 1995, MNRAS, 276, 625 

\bibitem[Kaiser 1992]{kai92}Kaiser, N., 1992, ApJ 388, 272

\bibitem[Kaiser \& Squires 1993]{kai93} Kaiser, N. \& Squires, G., 1993, 
ApJ, 404, 441

\bibitem[Kaiser 1997]{kai97}Kaiser, N., 1997, astro-ph/9610120

\bibitem[Mart\'\i nez-Gonz\'alez, Sanz \& Cay\'on 1997]{kike97}
Mart\'\i nez-Gonz\'alez, E., Sanz, J.L.,  \& Cay\'on, L. 1997, ApJ, 
to appear in July 20th issue (astro-ph/9702231)

\bibitem[Mould et al. 1994]{mou94}
Mould, J., Blandford, R., Villumsen, J., Brainerd, T., Smail, I., Small, T.
and Kells, W. 1994, MNRAS, 271, 31

\bibitem[Narayan \& Bartelmann 1996]{nar96} 
Narayan, R., \& Bartelmann, M. 1996, Proceedings of the 1995 
Jerusalem Winter School, eds. A. Dekel and J.P. Ostriker
 
\bibitem[Peacock \& Dodds 1996]{} Peacock, J.A. \& Dodds, S.J. 1996, MNRAS,
280, L19.

\bibitem[Schneider et al. 1992]{sch92} Schneider, P., Ehlers, J., \&
Falco, E.E. 1992, Gravitational Lenses (Heidelberg: Springer)

\bibitem[Seitz \& Schneider 1995]{sei95} Seitz, S., \& Schneider,
P. 1995, A\&A, 302, 9

\bibitem[Seljak 1996]{sel98}Seljak, U., 1996, ApJ, 463, 1

\bibitem[Wu \& Han  1995]{wuha95} Wu, X.P., Han, J., 1995, MNRAS, 272, 705

\bibitem[Wu 1996]{wu96} Wu, X.P. 1996, Fundamentals of Cosmic Physics,
in press

\bibitem[Viana \& Liddle 1996]{} Viana, P.T.P. \& Liddle, A.R. 1996, MNRAS, 
281, 323.

\bibitem[Villumsen 1995a]{vil95a}Villumsen, J. V. 1995, astro-ph/9507007

\bibitem[Villumsen 1995b]{vil95b}Villumsen, J. V. 1995, astro-ph/9512001

\bibitem[Villumsen 1996]{vil96} Villumsen, J. V. 1996, MNRAS, 281, 369

\bibitem[White et al. 1993]{} White, S.D.M., Efstathiou, G. \& Frenk, C.S.
1993, MNRAS, 262, 1023.

\end{thebibliography}
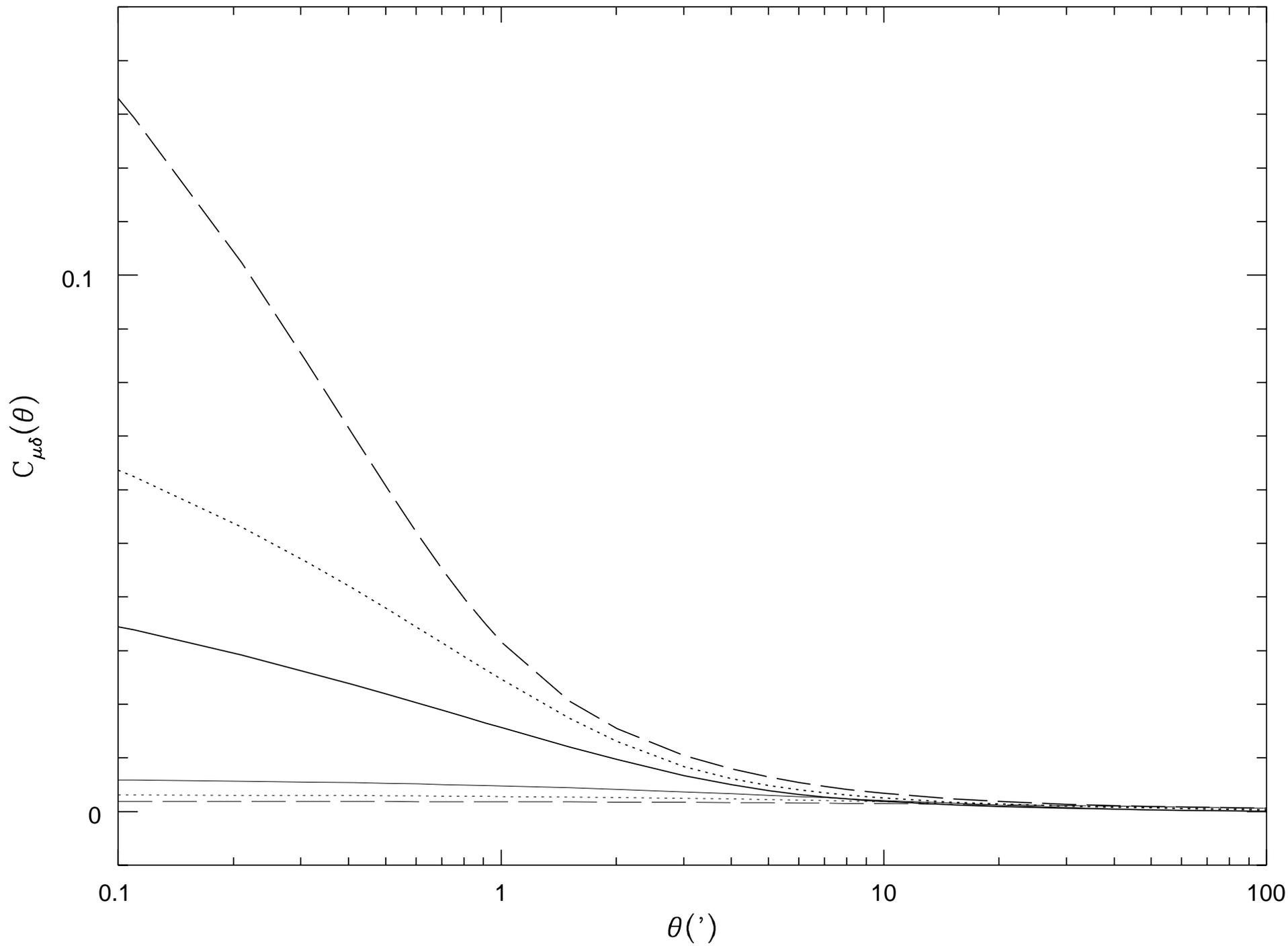
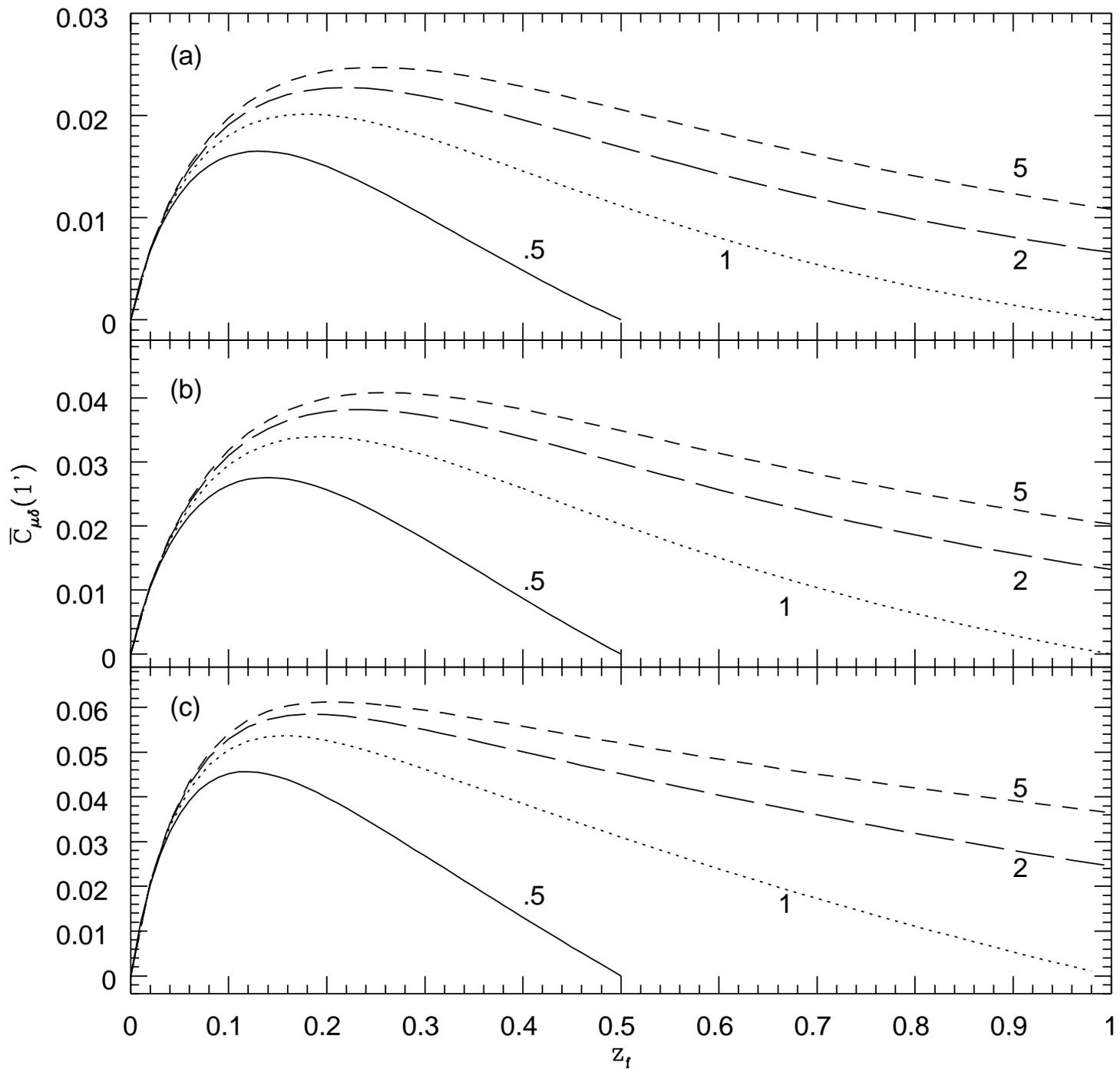
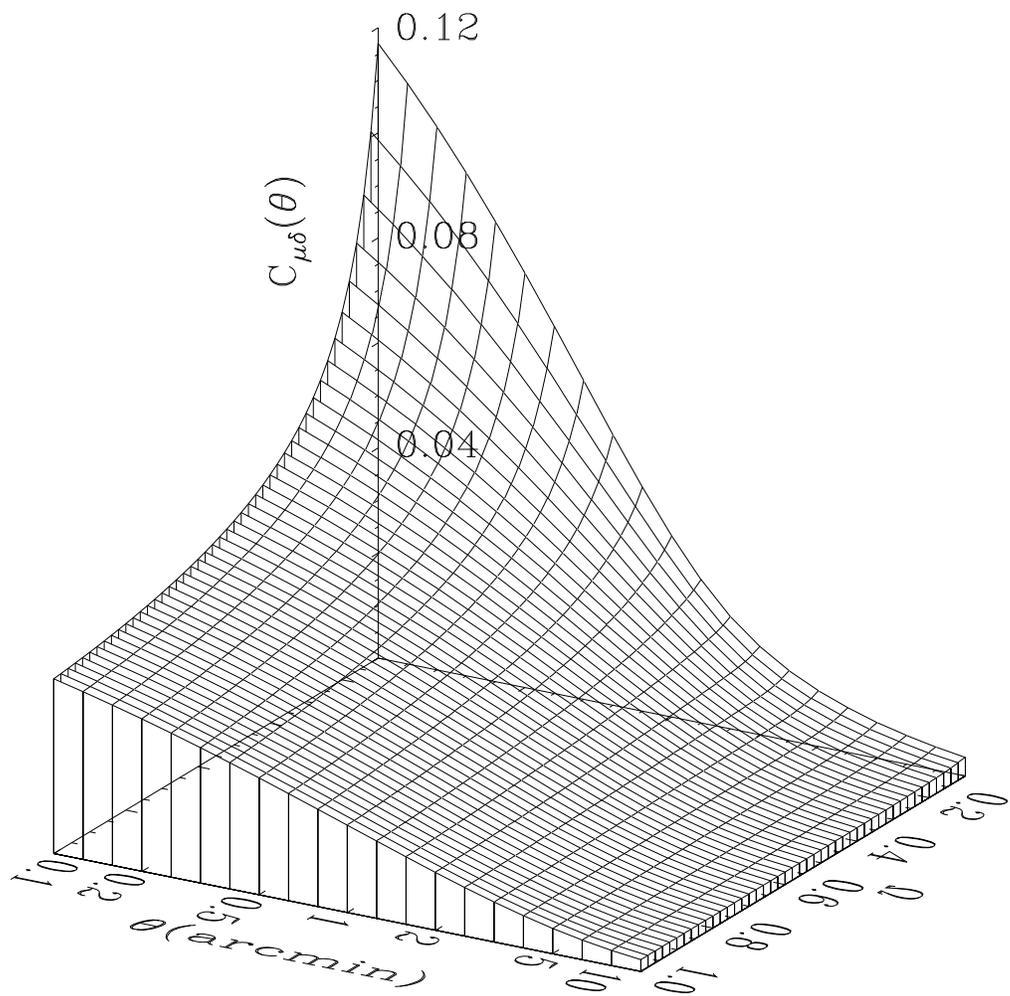
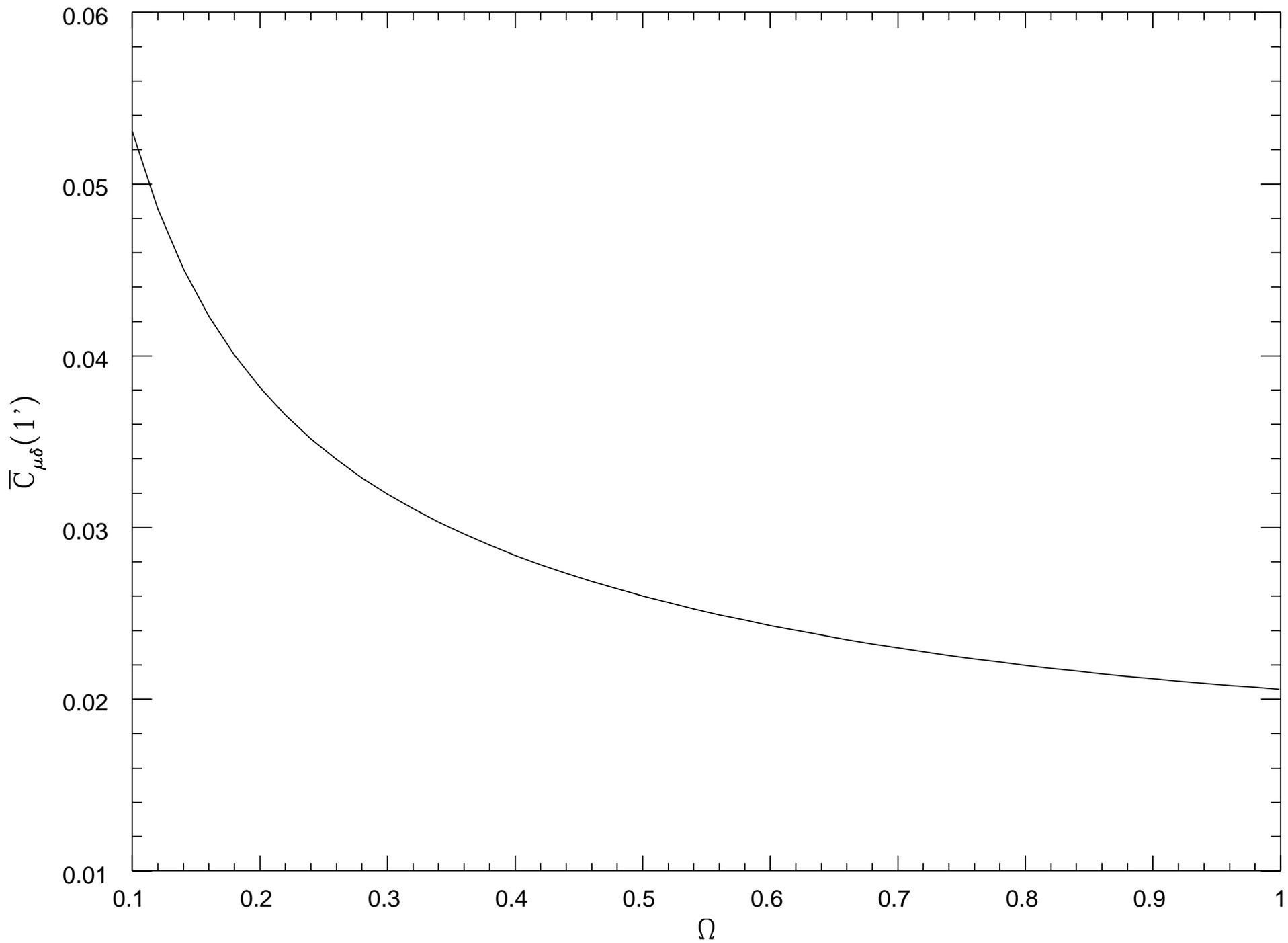
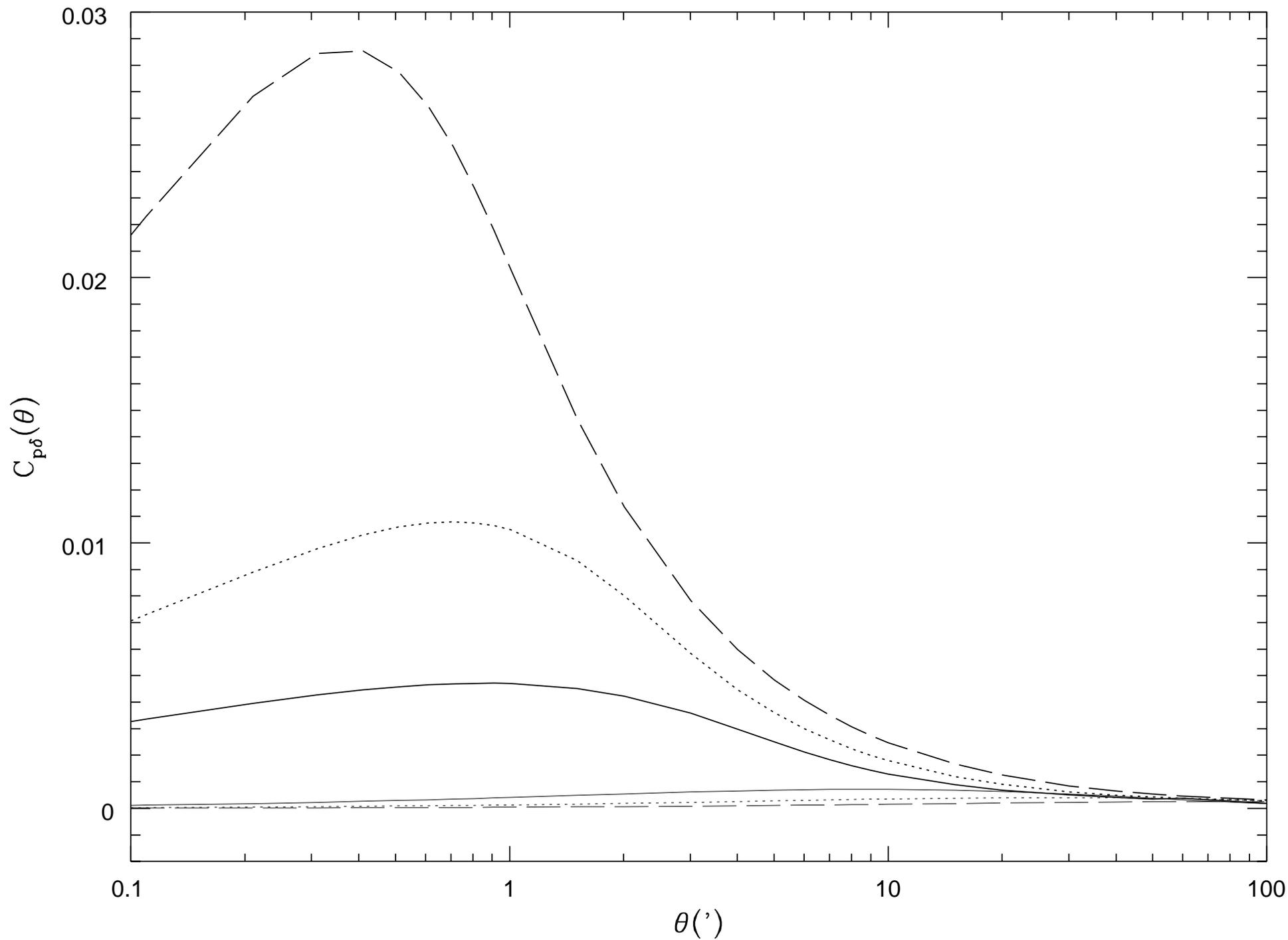
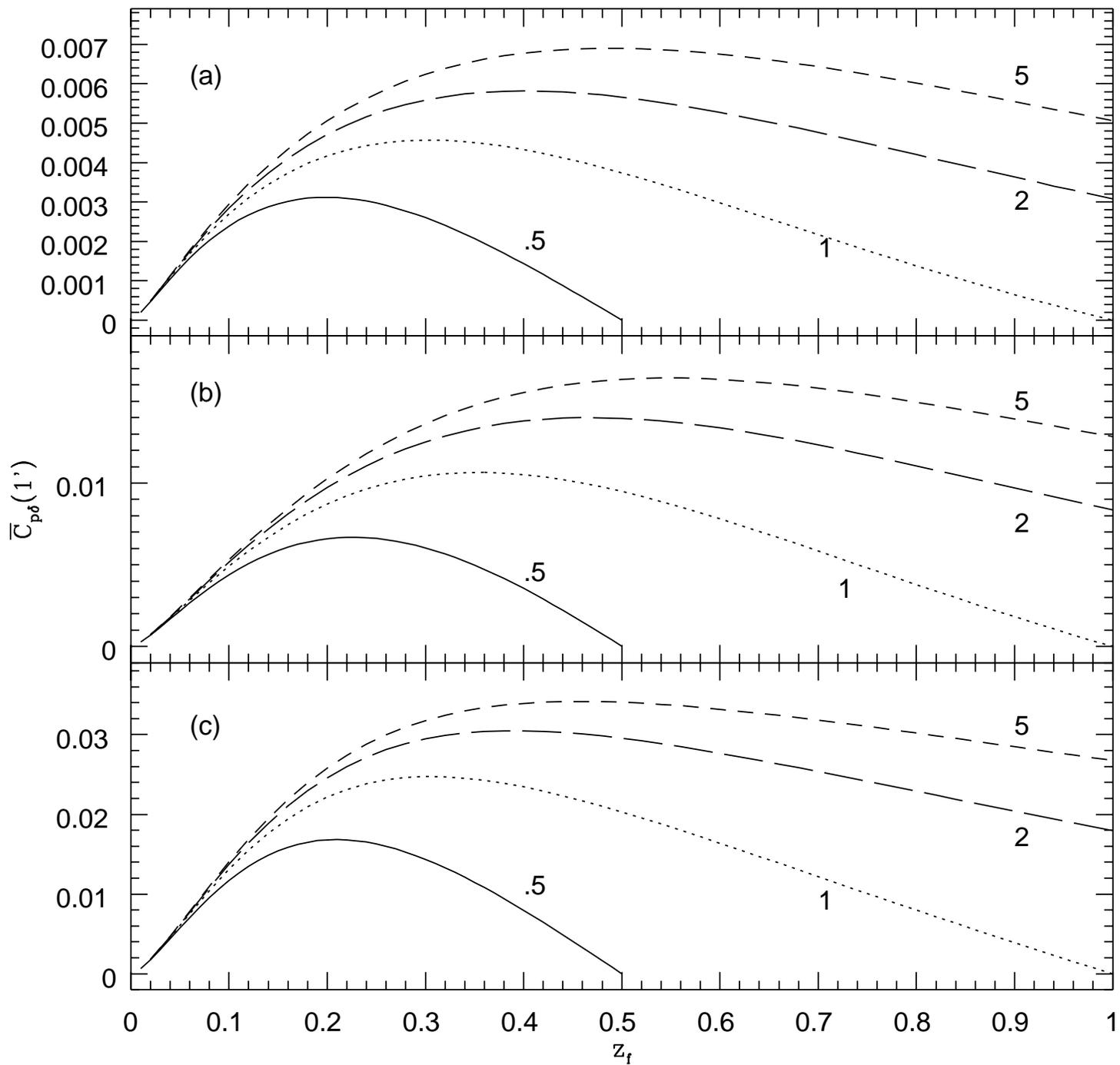
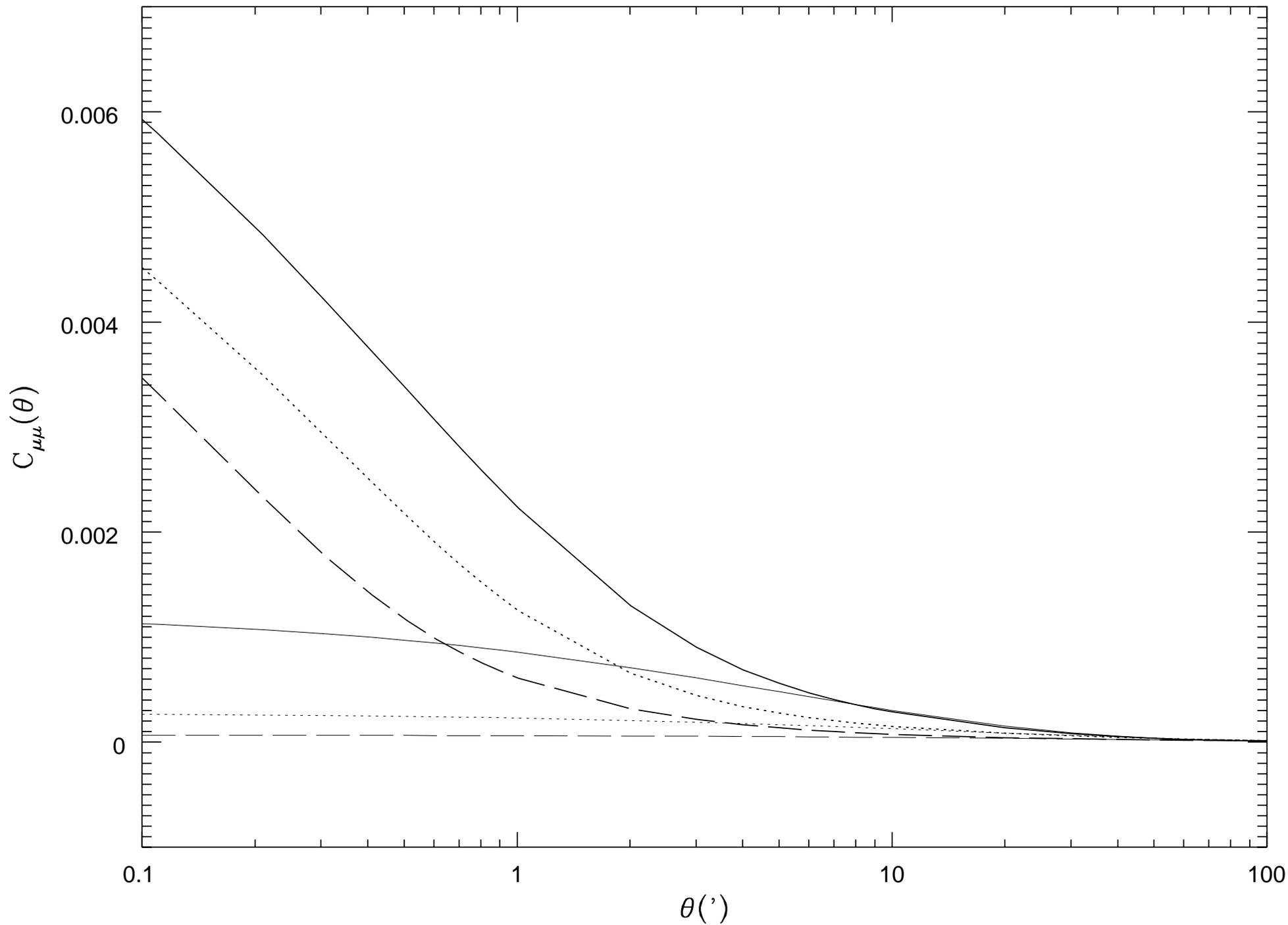
\end{document}